\documentclass[aps,preprint,prl,%
nofootinbib,
preprintnumbers,amsmath,amssymb,superscriptaddress,letter,longbibliography]{revtex4-1}
\usepackage[utf8]{inputenc}
\usepackage{graphicx}
\usepackage{subfigure}
\usepackage{color}
\usepackage{changes}
\usepackage{changes}
\renewcommand{\vec}[1]{\boldsymbol{#1}}
\usepackage[english]{babel}
\usepackage[pdftex]{hyperref}
\hypersetup{ 
pdfstartview=FitV,
colorlinks=true, 
linkcolor=black, 
citecolor=blue, 
filecolor=black, 
urlcolor=magenta,
}

\newcommand{\bea}{\begin{eqnarray}}
\newcommand{\eea}{\end{eqnarray}}

\graphicspath{{Figures/}}

\newcommand{\BG}[1]{{{#1}}} 
\newcommand{\EM}[1]{{{#1}}} 

\begin{document}

\preprint{CPHT-RR100.122019}

\title{Normal charge densities in quantum critical superfluids}


\author{Blaise {\sc Gout\'eraux}}\email[]{blaise.gouteraux@polytechnique.edu
}

\author{Eric {\sc Mefford}}\email[]{eric.mefford@polytechnique.edu
}

\affiliation{CPHT, CNRS, \'Ecole Polytechnique, IP Paris, F-91128 Palaiseau, France}

\begin{abstract}
The normal density of a translation-invariant superfluid often vanishes at zero temperature, as is observed in superfluid Helium and conventional superconductors described by BCS theory. Here we show that this need not be the case. We investigate the normal density in models of quantum critical superfluids using gauge-gravity duality. Models with an emergent infrared Lorentz symmetry lead to a vanishing normal density. On the other hand, models which break the isotropy between time and space may enjoy a non-vanishing normal density, depending on the spectrum of irrelevant deformations around the underlying quantum critical groundstate. Our results may shed light on recent measurements of the superfluid density and low energy spectral weight in superconducting overdoped cuprates.
\end{abstract}

\date{\today} 
\maketitle
\section{Introduction} Much of traditional superfluid and BEC superconductor phenomenology can be explained by Landau and Tisza's simple two-fluid hydrodynamical model \cite{Landau, Tisza} and its relativistic generalizations \cite{Khalatnikov1, Khalatnikov2, Khalatnikov3, Khalatnikov4, Israel1, Israel2,Carter:1995if}.
The Landau-Tisza model describes the superfluid as a mixture of two fluid components, the normal, \BG{dissipative} state with charge density $\rho_n$ and velocity $u_\mu$ and the \BG{dissipationless} superfluid with charge density $\rho_s$ and flow velocity $v_\mu$. The total charge density is the sum of both components, $\rho = \rho_n + \rho_s$. Experiments and theoretical calcuations in $^3$He, $^4$He, cold atom experiments, and conventional BCS superconductors all lead to the result that the system becomes entirely superfluid at zero temperature, ie that the normal density vanishes:
\begin{align}
\label{zerorhon}
\rho_{n}^{(0)} \equiv \lim_{T\to 0 } \rho_n = 0.
\end{align}
In \cite{Leggett,Leggett2}, two arguments were given to account for \eqref{zerorhon}. One argument used only the superfluid hydrodynamic description, the other assumed a weakly-coupled, Galilean, time-reversal invariant, single species superfluid.

The expectation \eqref{zerorhon} was called into question by recent experimental reports of anomalously low superfluid densities in overdoped high-$T_c$ superconductors \cite{Bozovic1} (see \cite{Zaanen2016,Kivelson2016} for commentary). Subsequent spectroscopic studies \cite{Bozovic2} also revealed very little loss of low energy spectral weight at low temperatures in the superconducting phase, suggesting a nonvanishing $\rho_{n}^{(0)}$. While the authors of \cite{LeeHone1,LeeHone2,LeeHone3} attributed this to disorder effects that can be captured in the so-called `dirty BCS' theory, fitting the experimental data relies on an ad hoc renormalization of the Drude weight \cite{LeeHone2}. Thus, no theoretical consensus has been reached on the experimental findings of \cite{Bozovic1,Bozovic2}, see also \cite{Bozovic3,Bozovic4}.

In this work, we tackle this question by combining methods using superfluid hydrodynamics and gauge-gravity duality. We review translation and time-reversal invariant superfluid hydrodynamics and show that the hydrodynamic equations are not enough to conclude that \eqref{zerorhon} is true. Instead, determining $\rho_{n}^{(0)}$ requires knowledge of the infrared (IR) equation of state. Using holographic models with quantum critical dynamics in the infrared as examples, we show \eqref{zerorhon} holds for strongly-coupled superfluids with an emergent Lorentz symmetry, in agreement with \cite{Leggett,Leggett2}. This is also consistent with the superfluid effective field theory discussed in \cite{Son:2002zn,Nicolis:2011cs,Delacretaz:2019brr}. On the other hand, we find that non-relativistic quantum critical systems with dynamical critical exponent $z>1$ can have $\rho_n^{(0)}\neq 0$. Hence, we conclude that a non-vanishing $\rho_n^{(0)}$ is not a result of the breakdown of the two-fluid model but rather a result of the quantum critical nature of the IR of these superfluids. Even after explicitly breaking translations, we show this conclusion does not change. These findings may suggest the anomalously low superfluid density and suppression of spectral weight observed in \cite{Bozovic1, Bozovic2} might be a consequence of the quantum critical properties of the superconducting phase of overdoped cuprates.

\BG{On a more formal level, our results indicate that the quantum effective action of Lifshitz superfluids differs significantly from that of their Lorentzian cousins \cite{Son:2002zn}, which opens exciting perspectives for future research on the theory of superfluids.}

\section{$\rho_n^{(0)}$ in superfluid hydrodynamics} 
In this section, we review relativistic, charged, superfluid hydrodynamics, following \cite{Herzog:2008he}. For our purposes, it is sufficient to work at the non-dissipative level.  Our results apply to any theory with translation invariance, including Galilean invariant theories. Relativistic symmetry leads to simpler notation and aligns nicely with our holographic example. A more thorough derivation can be found in Appendix \hyperref[apphydro]{A}. 

The system is described by the following equations (setting the speed of light $c=1$)
\begin{align}
\label{hydroequations_one}
\partial_\mu T^{\mu\nu} &= 0 ,\quad&\partial_\mu j^{\mu} =0,\nonumber\\
u^\mu\partial_\mu\varphi + \mu &= 0,\quad&\partial_\mu(su^\mu) = 0.
\end{align}
The first line expresses the local conservation laws: the conservation of the fluid stress tensor and the conservation of the $U(1)$ symmetry current, respectively. The last line states the constraints from gauge invariance and thermodynamics; respectively, a ``Josephson equation" which relates the time component of the background gauge field to the phase of the superfluid, $\varphi$, and the statement that in equilibrium, the entropy density is conserved. \BG{For simplicity, we have turned off external sources, which in particular corresponds to a choice of gauge $A_\mu=0$ for the external gauge field. The conclusions of this work are independent of the choice of gauge, see Appendix \hyperref[apphydro]{A}.}


Hydrodynamics states that these equations can be solved in terms of a derivative expansion of local thermodynamic variables and the fluid velocity. At non-dissipative order, \BG{thermodynamics of the equilibrium state fixes}
\begin{align}
\label{Eq:currents}
T^{\mu\nu} &= (\epsilon_n + P)u^\mu u^\nu + P\eta^{\mu\nu} +\frac{\rho_s}{\mu} \partial^\mu\varphi \partial^\nu\varphi\nonumber\\
j^\mu &= \rho_n u^\mu + \frac{\rho_s}{\mu}\partial^\mu\varphi.
\end{align}
The total charge density is the sum of the normal, $\rho_n$, and superfluid, $\rho_s$, densities. \EM{The distinction between $\rho_n$ and $\rho_s$ follows from the expectation that $u^\mu$ is the velocity of entropy flow which is carried purely by the normal component.} The normal energy density, $\epsilon_n$, and pressure, $P$, satisfy the Smarr and Gibbs relations,
\begin{align}
\label{Eq:thermoequilibrium}
&\epsilon_n + P = Ts+\rho_n\mu\,, \nonumber\\
&dP = sdT + \rho d\mu - \frac{\rho_s}{2\mu} d(\partial_\nu\varphi\partial^\nu\varphi + \mu^2)\,.
\end{align}

We perturb about equilibrium, writing $T=T_0 +\delta T$, $\mu = \mu_0 + \delta\mu $, $u^\mu= (1,\delta u^i)^\mu$, $\varphi = -\mu_0 t+\delta\varphi$. The fluctuation equations can be massaged into the form
\begin{align}
\label{fluctuationequation}
0=(\mu\rho_n + sT)\partial_t\delta u_i + \rho_s\partial_t\partial_i\delta\varphi + s\partial_i T + \rho\partial_i\delta\mu.
\end{align}
If $s\to 0$ as $T\to 0$, consistency of this equation requires
\begin{align}
\label{consistencyequation}
\rho_n^{(0)}\partial_t(\mu\delta u_i - \partial_i\delta\varphi) = 0.
\end{align}
If $\delta u_i$ and $\delta\varphi$ were allowed to fluctuate independently, we would conclude $\rho_n^{(0)} = 0$, as in \cite{Leggett,Leggett2}. 

 However, introducing an external source for $\varphi$ through $\delta H=\int d^2 x s_\varphi \cdot \partial\varphi $ leads to  $s_\varphi^i = \rho_s(\partial^i\varphi-\mu u^i)$, see Appendix \hyperref[apphydro]{A} and \cite{Valle:2007xx}. Setting the external source to zero,  the superfluid velocity $v^i\equiv\partial_i\varphi/\mu$ is aligned with the fluid velocity $u^i$ and equation \eqref{consistencyequation} is automatically satisfied. Therefore, consistent coupling of the hydrodynamics to external sources evades the conclusion that $\rho_n^{(0)}=0$.

The fluctuations lead to an electrical conductivity at non-dissipative order \cite{KadanoffandMartin},\footnote{\BG{In general, contact terms may affect the conductivities \cite{Kovtun:2012rj}, but this is not the case for the electric conductivity of interest here.}}
\begin{align}
\label{hydrodynamicconductivity1}
\sigma(\omega) = \frac{i}{\omega}G^R_{J^x J^x}(\omega,0) = \frac{i}{\omega}\left[\frac{\rho_n^2}{\mu\rho_n+sT} + \frac{\rho_s}{\mu}\right].
\end{align}
Importantly, $\lim_{T\to 0} \omega Im[\sigma]=\rho/\mu$, irrespective of whether $\rho_n^{(0)}=0$ or not. \BG{Here, as well as everywhere in the rest of our work, we take the $\omega\to0$ limit before the $T\to0$ limit}. The Kramers-Kronig relations require that $Re[\sigma]$ also has a delta function as $\omega\to 0$ with the same weight. Eq.~\eqref{hydrodynamicconductivity1} applies equally well to superconductors with a dynamical gauge field, as the conductivity is measured with respect to the total electric field, which relates it to the unscreened retarded Green's function. 

If we explicitly break translations weakly, the momentum relaxes slowly with an inverse lifetime $\Gamma$ and the conductivity becomes
\begin{align}
\label{hydrodynamicconductivity2}
\sigma(\omega) = \frac{\rho_n^2}{\mu\rho_n+sT}\frac{1}{\Gamma-i\omega} + \frac{\rho_s}{\mu}\left(\frac{i}{\omega}\right).
\end{align}
The imaginary pole is now proportional only to the superfluid density, though this says nothing about $\rho_n^{(0)}$. Importantly, there are no inconsistencies if $\rho_n^{(0)} = 0$ when translations are broken, as we will demonstrate.

\section{Holographic quantum critical superfluids} Holography relates the low energy dynamics of a finite temperature strongly interacting gauge theory with a large number of colors in $d+1$ spacetime dimensions to the dynamics of a classical gravitational system in $d+2$ dimensions with a black hole \cite{Maldacena:1997re,Witten:1998qj}. While explicit examples are known from string theory which fix the action of the gravitational theory, applied holography posits that a consistent set of a small number of fields, such as scalars and $U(1)$ gauge fields, coupled to gravity in $(d+2)$ anti-de Sitter spacetime is able to capture the universal low energy dynamics of a large number of strongly interacting quantum systems near a quantum critical point or phase \cite{Hartnoll:2016apf}. 

In particular, these quantum critical theories should be characterized by the dependence of correlation functions on certain universal exponents, for instance the dynamical critical exponent, $z$, the hyperscaling violation parameter $\theta$, and the spatial dimension $d$. Holographically, these exponents are captured by an extremal (zero temperature) horizon of the form \cite{Charmousis:2010zz,Davison:2018nxm}
\begin{align}
\label{IRmetric}
ds^2 = r^{2\frac{\theta}{d}-2}\left[-L_t^2 \frac{dt^2}{r^{2z-2}} + \tilde{L}^2dr^2 + L_x^2d\vec{x}^2\right]
\end{align}
where the horizon is at $r\to \infty$ when $z\geq1$. The radial coordinate $r$ functions as a renormalization scale so that under scale transformations,
\begin{align}
(r,x^i)\to \lambda (r,x^i)\quad t\to \lambda^{z} t \Rightarrow ds^2 \to \lambda^{2\frac{\theta}{d}}ds^2.
\end{align}
This implies that the thermodynamic parameters have dimension $[T] = z$ and $[s] = d-\theta$ and $s\sim T^{\frac{d-\theta}{z}}$. 

A very general gravitational model which can lead to these extremal solutions is the following, \cite{Adam:2012mw},
\begin{align}
\label{holographicaction}
S = \int d^{d+2}x&\sqrt{-g}\Biggl[R-\frac{Z(\phi)}{4}F^2 - |D\eta|^2 - \frac{1}{2}(\partial\phi)^2- V(\phi,|\eta|) - \frac{Y(\phi)}{2}\sum_{i=1}^{d}(\partial\psi_i)^2\Biggr].
\end{align}
Here, $A_M$ is a $U(1)$ gauge field with field strength $F_{MN} = \partial_MA_N-\partial_NA_M$. The field $\eta$ is a complex scalar with $U(1)$ charge $Q$ and covariant derivative $D\eta = (\partial_M - iQA_M)\eta$. When $|\eta|\neq 0$, the $U(1)$ symmetry is broken and the dual theory can be thought of as a superfluid \cite{Hartnoll:2008kx, Hartnoll:2008vx}. The field $\phi$ is a neutral scalar called the dilaton which has a source on the boundary $\phi_s$. The fields $\psi_I$ are chosen to have linear dependence on the spatial dimensions, $\psi_i = mx^j\delta_{ij}$ so that when $Y\neq 0$, they explicitly break translation but not rotation invariance \cite{Andrade:2013gsa}. The gauge field is chosen only to have a background time component whose value at the boundary of $AdS$ sets the chemical potential, $\mu$, which sources a charge density, $\rho$. We have set $16\pi G = 1$. See Appendix \hyperref[app:model]{B} for further details.

The relativistic invariant equations of superfluidity described above were shown to hold in holographic models where the transport coefficients can be derived from the gravitational dual to the boundary fluid \cite{Herzog:2008he, Herzog:2009md, Sonner:2010yx, Herzog:2011ec, Bhattacharya:2011eea,Bhattacharya:2011tra}. Though the early holographic models focused on the original holographic superfluid \cite{Gubser:2008px, Hartnoll:2008kx, Hartnoll:2008vx}, the bulk action can be generalized as in \eqref{holographicaction} to include running couplings and bounded scalar potentials \cite{Gubser:2009cg, Horowitz:2009ij, Adam:2012mw, Gouteraux:2012yr}. We find that the two-fluid hydrodynamic model still works well in describing these models. 

The solutions \eqref{IRmetric} are found for potentials which behave in the IR as 
\begin{align}
Y(\phi\to\infty)&\to Y_0e^{\lambda\phi},\quad V(\phi,|\eta|)\to V_0e^{-\delta\phi},\nonumber\\
Z(\phi\to\infty)&\to Z_0e^{\gamma\phi}, \quad \phi = \kappa\ln\left(r\right).
\end{align}

The gauge field and translation breaking scalars can be engineered to be marginal or irrelevant deformations of the IR critical phase, \cite{Gouteraux:2014hca,Davison:2018nxm}. We will be concerned with phases where the charged scalar is irrelevant in the IR, taking the asymptotic value $\eta_0$ \cite{Adam:2012mw}. This implies that the scaling exponents are the same in the superfluid as in the normal phase, so that many of the scaling properties at low temperature are inherited from the normal phase.

\section{Normal densities in holographic superfluids} To find the normal density (see also \cite{Herzog:2009md}), we perturb our system by turning on a spatially homogeneous infinitesimal external electric field in the $x$-direction, $E_xe^{-i\omega t}$, sourcing both an electric and a momentum current, see Appendix \hyperref[app:model]{B}. \BG{As $\omega\to 0$, the equation for the momentum current enforces $\langle T_{tx} \rangle = -\rho \delta\xi_x$,
where} \EM{ $\delta\xi_x= \partial_x\varphi - A_x$} \BG{is gauge-invariant, and given by the electric field $\delta\xi_x = E_x/(i\omega)$ in a gauge where $\varphi=0$, which is the gauge we work in. This response requires that $\mu\delta u_x= \delta\xi_x$, see Appendix \hyperref[app:conductivity]{C}. }

In the companion {paper \cite{normaldensitypaper2}, we explore transport in the superfluid phases of the holographic model \eqref{holographicaction} for general potentials in greater detail. Here, for illustrative purposes, we present an explicit example in $d=2$ that leads to $\rho_n^{(0)}\neq 0$ and one that leads to $\rho_{n}^{(0)}=0$, including when translations are broken. Specifically, we use the model of \cite{Adam:2012mw} with
\begin{align}
Z &= e^{\frac{\phi}{\sqrt{3}}}, \;\;V= -6\cosh\left(\phi/\sqrt{3}\right) - 2|\eta|^2 +|\eta|^4, \;\; Y=0.
\end{align}
Upon varying the dilaton source, this model has two IR phases characterized by critical exponents,
\begin{align}
\left(z,\theta,\frac{z}{\theta}\right) =  (+\infty,-\infty,-1)\quad \text{or} \quad (z,\theta) = (1,-1).
\end{align}
{In the first case, we first need to redefine $r\mapsto r^{1/z}$ before sending $z\to+\infty$ in \eqref{IRmetric}.
The IR behavior of $\phi$ in the two phases is $\phi =  \pm\sqrt{3}\ln\left(r\right)$.
In the first case, $Z(\phi)$ diverges and leads to a finite electric flux, $\rho_{in}^{(0)}$, from the extremal horizon, suggesting a ``fractionalization" of charged degrees of freedom into a subset confined in the condensate and subset deconfined in the thermal bath hidden by the horizon \cite{Hartnoll:2011fn,Hartnoll:2011pp}. In the second case, $Z(\phi)\to 0$ causing the flux to vanish in the IR and all charged degrees of freedom are confined into the charged condensate in a ``cohesive" phase. 

These two cases can also be distinguished by the vanishing of $\rho^{(0)}_n$ in the cohesive phase and non-vanishing of $\rho_n^{(0)}$ in the fractionalized phase. We emphasize that, despite their apparent similarities, $\rho_n^{(0)}$ and $\rho_{in}^{(0)}$ are not immediately related. The first is a quantity defined in the two-fluid hydrodynamic model while the second is a microscopic measurement of the uncondensed degrees of freedom. This is analogous to BEC superconductivity where not all electrons condense into Cooper pairs, yet $\rho_n\to 0$ \cite{altland_simons_2010}. In fact, in \cite{normaldensitypaper2}, we discuss pure Lifshitz superfluid solutions \cite{Gubser:2009cg,Horowitz:2009ij} in which $\rho_{in}^{(0)}$ vanishes while $\rho_n^{(0)}$ does not, for sufficiently large $z$.

\begin{figure}[t]
\centering
\includegraphics[width=.45\columnwidth]{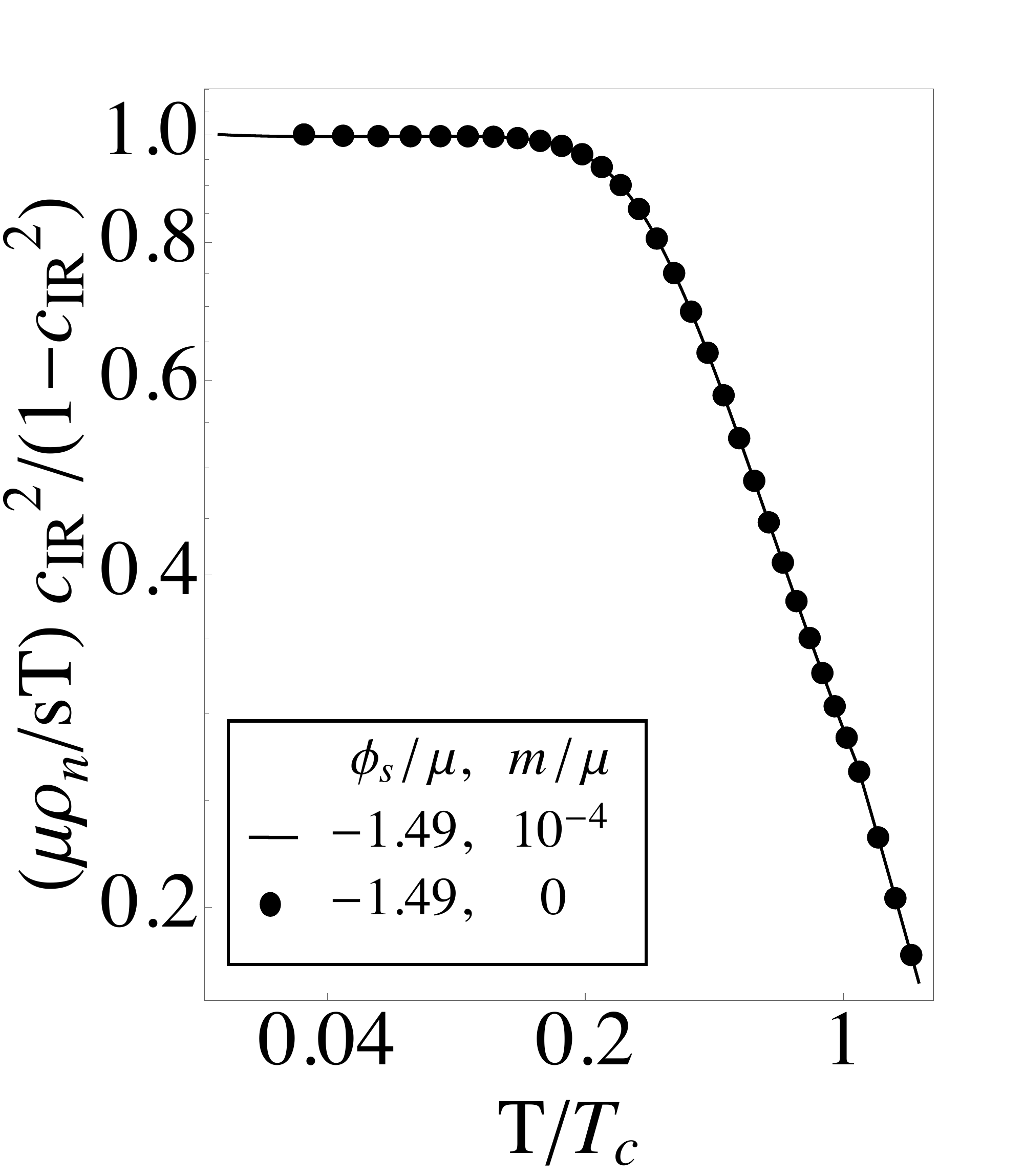}
\includegraphics[width=.45\columnwidth]{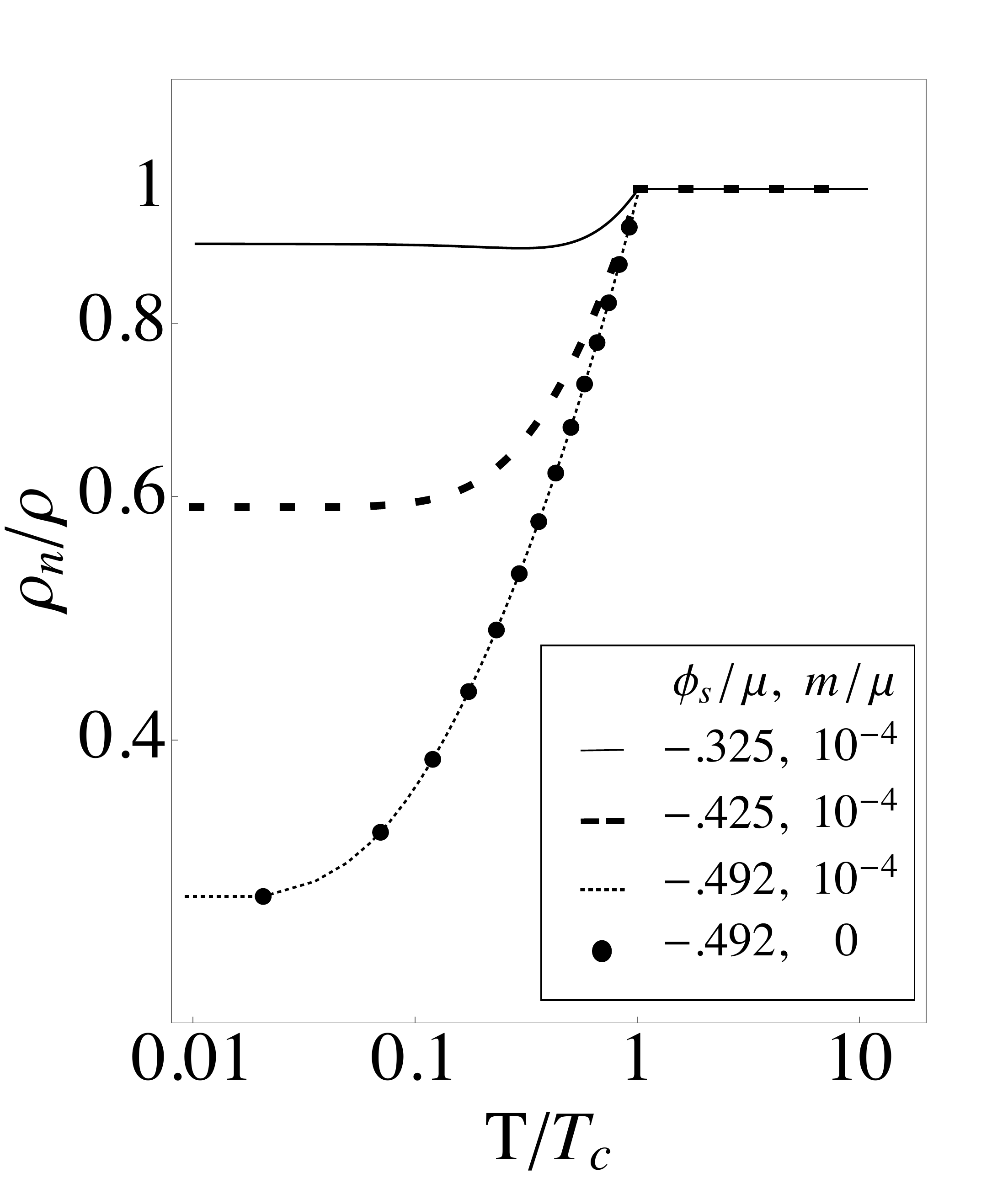}
\caption{Temperature dependence of $\rho_n/\rho$ in the cohesive (left) and fractionalized (right) phases. }
\label{fig:1}
\end{figure}

After solving for the bulk $\delta\xi_x(r)$, we combine \eqref{hydrodynamicconductivity1} with the knowledge of the total background charge density $\rho=\rho_s+\rho_n$ to extract both $\rho_n$ and $\rho_s$. Our numerical results are shown in Fig.\ref{fig:1}. In \cite{normaldensitypaper2}, we show analytically that
\begin{eqnarray}
\label{rhonbehaviorfrac}
\text{fractionalized: }&\rho_n&\simeq\;\;\frac{\rho_n^{(0)}}{\mu^2} +\# T^{1-\frac\theta{z}}+...\;\; \\
\label{rhonbehaviorcoh}
\text{cohesive: }&\rho_n&\simeq \frac{1-c_{IR}^2}{c_{IR}^2}\, \frac{s T}{\mu} +...\sim  T^{\frac{(2+z-\theta)}z}\;\;
\end{eqnarray}
where $\rho_n^{(0)}$ depends on UV parameters, for instance, the source, $\phi_s$, and $c_{IR} \equiv L_t/L_x\,r_h^{1-z}\sim T^{1-1/z}$ is the lightcone velocity in the IR. The $...$ indicate terms from more irrelevant deformations of the IR geometry.
Interestingly, the leading order temperature dependencies of the normal density behave as power-laws with exponents determined by the underlying IR phase, characteristic of quantum critical systems. This is in contrast to BCS superconductivity, in which it is found that $\rho_n$ is exponentially suppressed \cite{Leggett}. On the other hand, in $^4$He, the normal (mass) density is controlled by phonons (\EM{the goldstones from the $U(1)$ breaking}) so that $\rho_n =  \frac{ (sT)}{c_p^2}+O(c_p^{-1})$ where the coefficient is the phonon speed of sound, $c_p$ \cite{Schmitt:2014eka}. This is identical to \eqref{rhonbehaviorcoh}, trading $c_{IR}\mapsto c_p$ and taking the limit $c_p\ll 1$.

\EM{In \cite{normaldensitypaper2}, we find that $\rho_{n}^{(0)}$ depends on the competition between two terms proportional to $sT$ and $c_{IR}^2$, respectively.\footnote{\BG{The terms arise from the breaking of particle-hole symmetry and gauge invariance, respectively. Denoting the relative temperature dependence of these two terms by $T^\alpha$, we find a more general criteria $\alpha < d+2-\theta-z \Rightarrow \rho_n^{(0)} = 0$. Here, we have $\alpha = 0$. In different models \cite{normaldensitypaper2} than the ones considered here, we can find fractionalized phases which have $\rho_{n}^{(0)}=0$, similar to {}$^4$He.}} If $c_{IR}^2$ dominates at low $T$, then $\rho_{n}^{(0)}\neq0$. Otherwise, $\rho_n^{(0)}=0$ and to leading order $\rho_n$ is given by \eqref{rhonbehaviorcoh}. This result is consistent with the relativistic superfluid effective field theory \cite{Delacretaz:2019brr}, but is also true for $z\neq1$.  For the quantum critical superfluids presented here, fractionalized phases ($\rho_{in}^{(0)}\neq 0$) always have $c_{IR}^2 > sT$ and hence $\rho_n^{(0)}\neq 0$, whereas for cohesive phases ($\rho_{in}^{(0)}=0$), this occurs for:
\begin{align}
\label{exponentcriteria}
\text{cohesive: }\qquad z<d+2-\theta\quad\Rightarrow\quad \rho_n^{(0)}=0.
\end{align}}
\BG{We observe that when \eqref{exponentcriteria} is violated, \eqref{rhonbehaviorcoh} would naively lead to a divergent $\rho_n^{(0)}$. Instead, as we have just explained, a more careful calculation leads to a finite $\rho_n^{(0)}\neq0$. }

Generically, many irrelevant deformations of the criticial IR geometry compete to drive the system toward the UV. In particular, while a universal deformation proportional to $sT$ always exists, dangerously irrelevant operators may control the temperature dependence of thermodynamic or transport observables \cite{Blake:2016jnn,Davison:2018ofp,Davison:2018nxm}. It is then remarkable that the criteria in \eqref{exponentcriteria} leads to the universal temperature dependence \eqref{rhonbehaviorcoh} for cohesive phases.

\begin{figure}[t]
\centering
\includegraphics[width=.45\columnwidth]{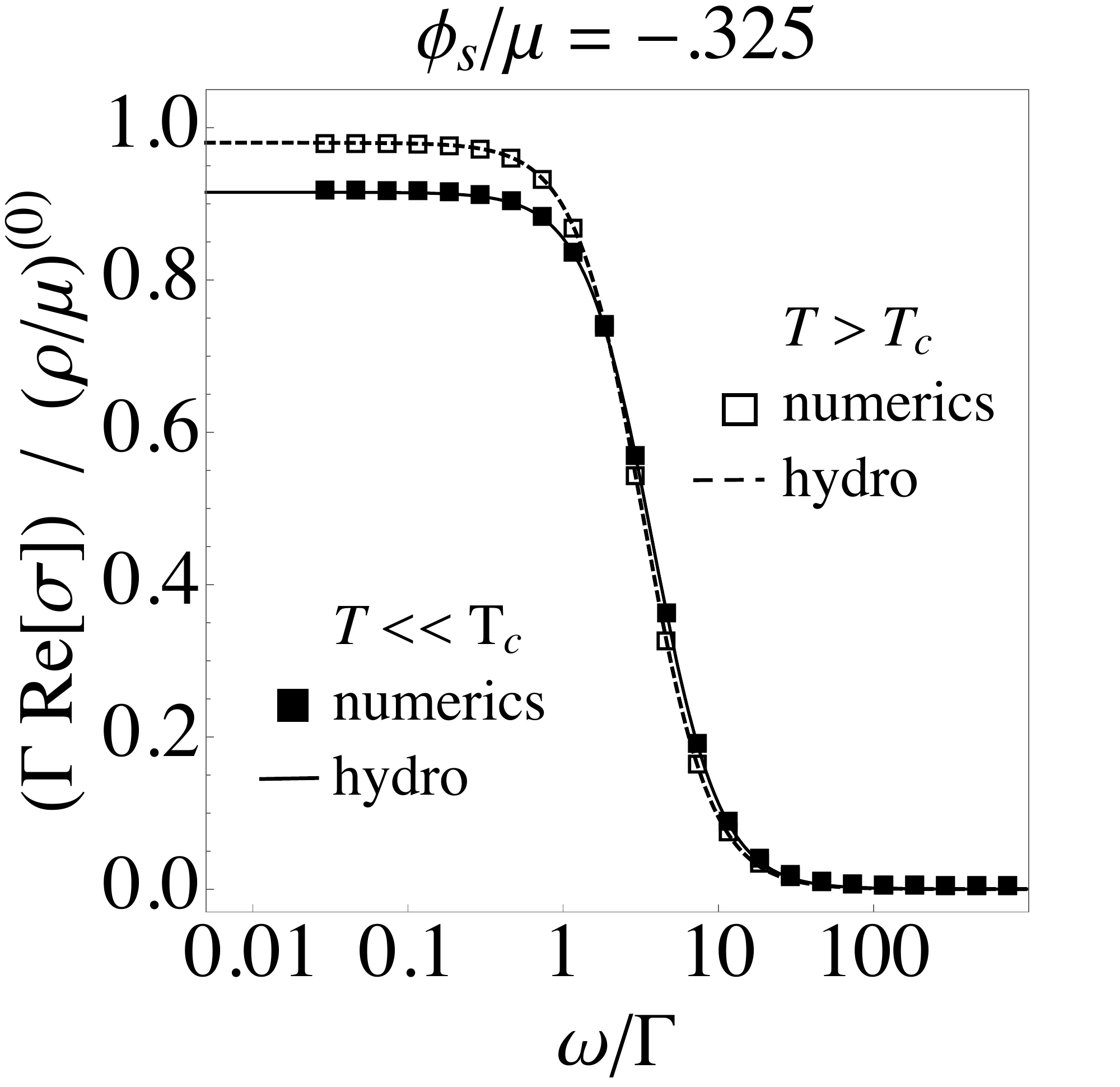}
\includegraphics[width=.45\columnwidth]{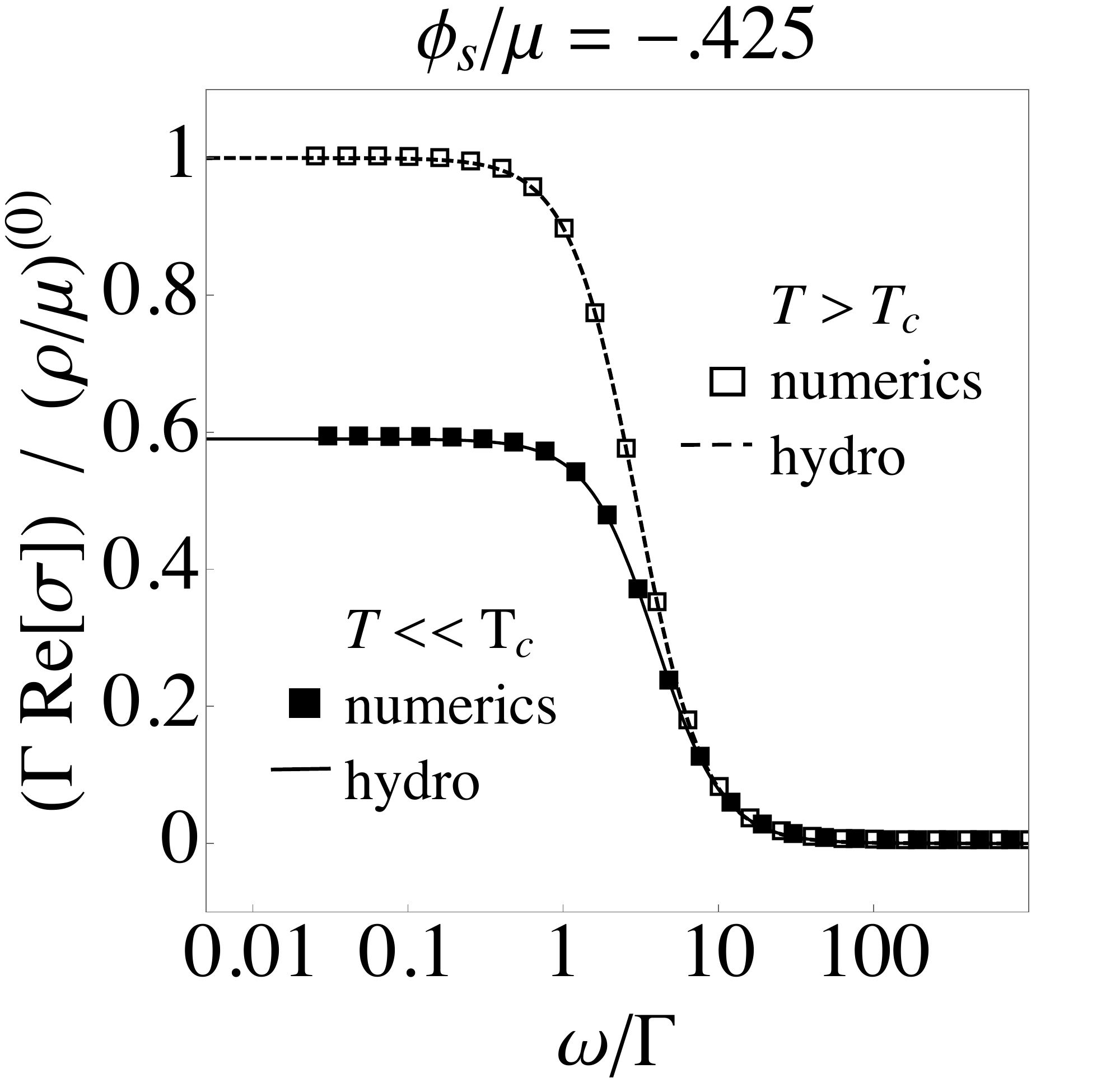}
\caption{Varying depletion of spectral weight in the real part of $\sigma(\omega)$ due to the superfluid in the presence of broken translations ($m/\mu=10^{-2}$). 
Lower depletion is correlated with a larger $\rho_n^{(0)}$. The curves are the real part of \eqref{hydrodynamicconductivity2}.}
\label{conductivity}
\end{figure}

As a final illustration that $\rho^{(0)}_n\neq 0$ is a signature of criticality rather than, for instance, disorder, we explicitly break translations in \eqref{holographicaction},
with $Y(\phi) = \exp\left(\mp\phi/\sqrt{3}\right)$,
where the minus sign is for fractionalized phases and the plus for cohesive phases. This choice ensures that translation breaking is sufficiently irrelevant to not destabilize the IR geometry. 
We omit the detailed accounting of gauge invariant fluctuations which can be found, for instance, in \cite{Donos:2013eha}. Due to the introduction of broken translations, we confirm $\lim_{\omega\to0}\omega\text{Im}[\sigma] = \rho_s/\mu$ (see also \cite{Ling:2014laa,Andrade:2014xca,Kim:2015dna}) as in \eqref{hydrodynamicconductivity2} and find $\Gamma = m^2sY(r_h)/(4\pi[\mu\rho_n+sT])$ as follows from \cite{Davison:2018nxm}, see figure \ref{conductivity}. Furthermore, the temperature dependence in Eqs.~(\ref{rhonbehaviorfrac},\ref{rhonbehaviorcoh}) does not change. In particular, translation symmetry breaking does not necessarily give rise to finite $\rho_n^{(0)}$ in the cohesive phase. Instead, $\rho_n^{(0)}$ is controlled by the underlying criticality.

\section{Low temperature behavior of hydrodynamic modes} Eq.~\eqref{rhonbehaviorcoh} has interesting consequences on the spectrum of hydrodynamic modes at low temperatures. The superfluid second sound velocity is given by \cite{Herzog:2009md,Herzog:2011ec} 
\begin{equation}
c_2^2=\left(\frac{s}{\rho}\right)^2\frac{\rho_s}{(sT+\mu\rho_n)(\partial[s/\rho]/\partial T)_\mu}\,.
\end{equation}
Using \eqref{rhonbehaviorcoh} and $s\sim T^{(d-\theta)/z}$, we find $c_2^2 = zc_{IR}^2/(d-\theta)$. This is the generalization of Landau's conjecture \cite{landaubook} to critical IR geometries. For fractionalized phases, on the other hand, we find $c_2^2\sim s T$, which decays parametrically faster with temperature than $c_{IR}$ when \eqref{exponentcriteria} holds. In both cases, the superfluid sound velocity vanishes at $T=0$. This is in marked contrast to the relativistic case $z=1$ and the superfluid effective field theory \cite{Delacretaz:2019brr}, which lead to a non-vanishing $T=0$ superfluid velocity. We expect the Goldstone mode should interpolate to a dispersion relation $\omega\sim k^z$ in the limit $T\ll k$. It would be interesting to work this out in our model.

Fourth sound is defined as the sound mode which propagates when the normal velocity vanishes \cite{landaubook}, given by
\begin{equation}
\label{fourthsound}
c_4^2=\frac{\rho_s}{\mu\left(\frac{\partial\rho}{\partial\mu}\right)_s}\simeq \frac{\rho_s}{d\rho}.
\end{equation}
In the second equality, we have used the low temperature behavior, $\rho\sim\mu^d$. Thus, fourth sound provides a direct measure of whether $\rho_n^{(0)}=0$, since then $c_4^2=1/d$. This result explains some observations reported in previous literature, \cite{Yarom:2009uq,Herzog:2009md}. In dirty superfluids with broken translations, only fourth sound survives. In particular, \eqref{fourthsound} matches the expressions in \cite{Davison:2016hno}. Thus, measuring superfluid sound in impure, quantum critical superfluids would give direct information on whether $\rho_n^{(0)}=0$ or not. 


\section{Discussion} In this manuscript, we have shown that a non-vanishing $\rho_n^{(0)}$ is consistent with the Landau-Tisza two-fluid model of superfluidity. This is because in the absence of external sources, fluctuations in the normal velocity are aligned with fluctuations in the superfluid velocity, $\mu\delta u_i = \partial_i\delta\varphi$. We illustrated this using a model of holographic superfluidity and showed that $\rho_n^{(0)}$ is controlled by the underlying quantum critical phase. Experimental evidence for a non-vanishing $\rho_n^{(0)}$ in the cuprates can be considered further evidence for the existence of an underlying quantum critical phase in those systems. It would be interesting to find more experimental examples of non-vanishing $\rho_n^{(0)}$, perhaps in cold atom experiments, which we expect would be a generic feature of quantum criticality.

\begin{figure}[t]
\centering
\includegraphics[width=.45\columnwidth]{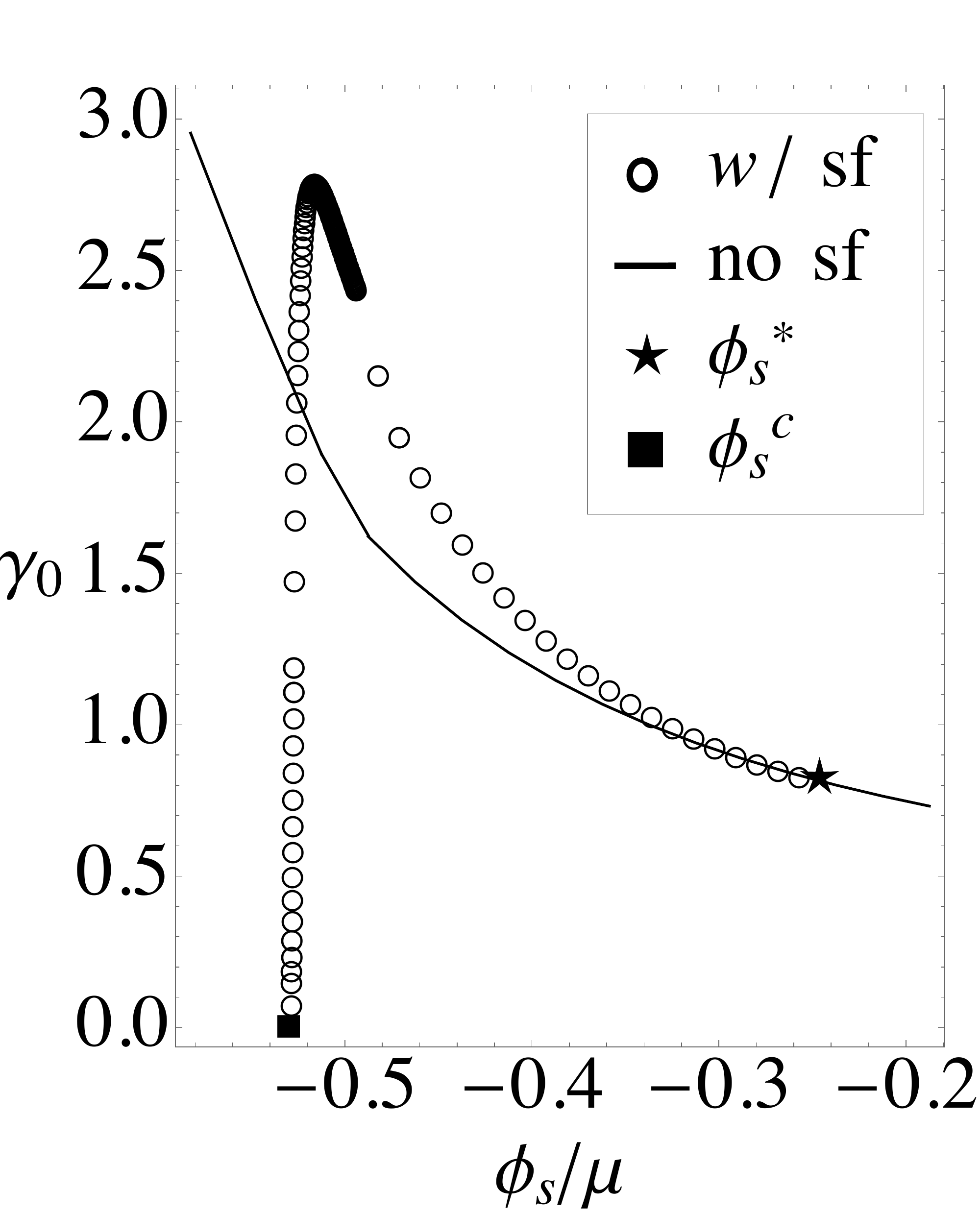}
\includegraphics[width=.45\columnwidth]{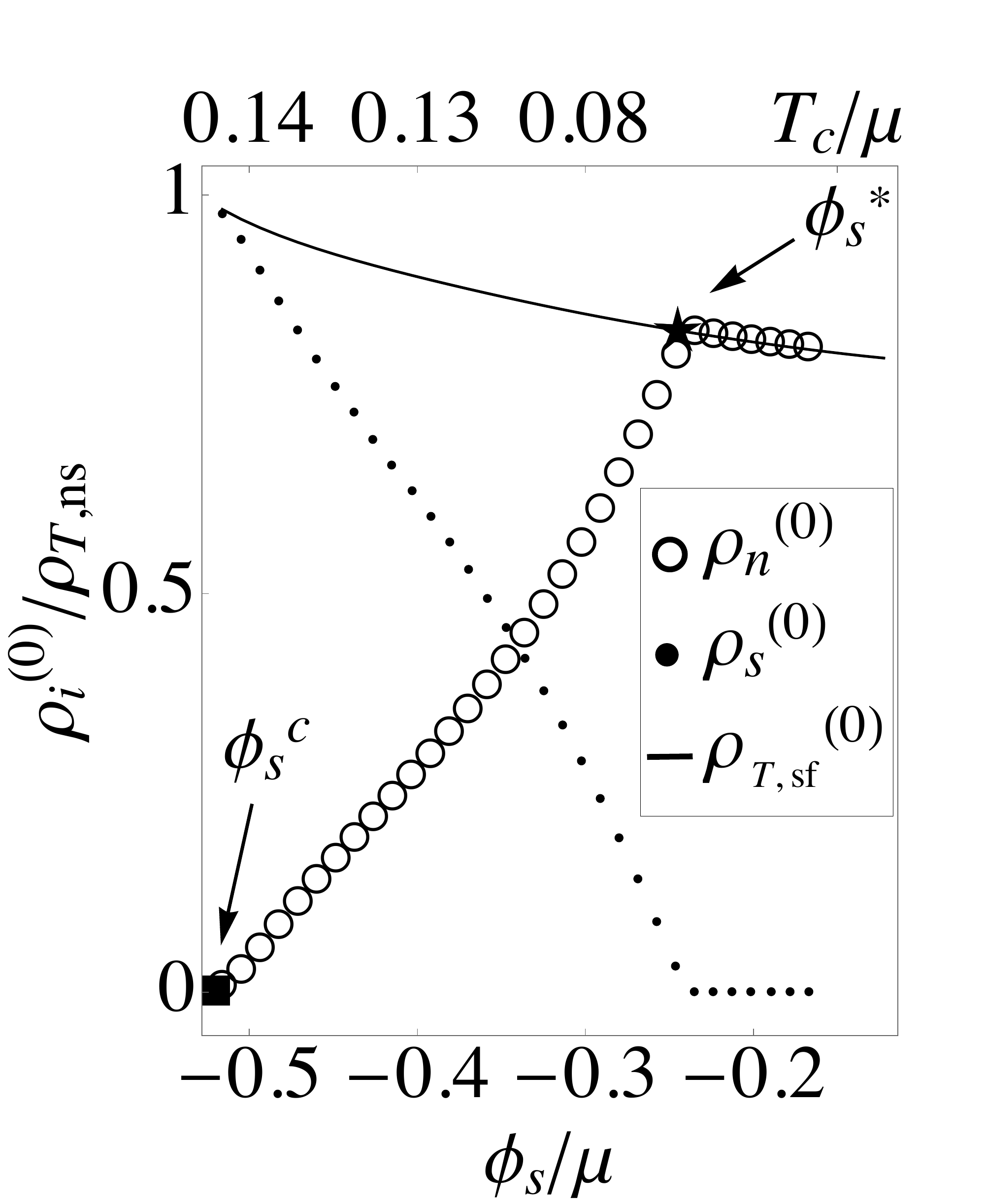}
\caption{Left: The linear specific heat slope, $\gamma_0$, as a function of the dilaton source, $\phi_s$, mimicking Fig.~(5) of \cite{HHWen1}. Right: The zero temperature densities as a function of $\phi_s$ compared to the total density with no superfluid, $\rho_{T,ns}$ mimicking Fig.~(2e) of \cite{Bozovic2}. Two special points are indicated: $\phi_s^c$ marks the phase transition between cohesive and fractionalized phases and $\phi_s^*$ marks the phase transition to a totally fractionalized phase with superfluidity completely suppressed, $\rho_s^{(0)}=0$. }
\label{densitiesandspecificheat}
\end{figure}

The holographic models discussed here exhibit further similarities to experimental observations in the cuprates. In overdoped La$_{2-x}$Sr$_x$CuO$_4$, \cite{HHWen1, HHWen2}, heat capacity measurements reveal a linear in temperature component at low temperatures, $c \sim \gamma_0 T + O(T^2)$. The coefficient, $\gamma_0$, is a measure of the density of normal charge carriers and exhibits strong doping dependence that is correlated with strong depletion of spectral weight in the optical conductivity \cite{Bozovic1,Bozovic2}. Interpreting a source for the dilaton, $\phi_s$, as a proxy for doping,\footnote{\BG{In using $\phi_s/\mu$ as a proxy for doping, we note that its variation triggers a quantum phase transition between a superfluid and normal phase. Similarly to doping, increasing $\phi_s/\mu$ increases the total charge density, $\rho_T^{(0)}$, as well as amplifies the effect of momentum breaking (increased disorder strength), see Appendix \hyperref[app:doping]{D}.}} our models exhibit the same behavior, illustrated in figures \ref{conductivity} and \ref{densitiesandspecificheat}. The rapid depletion arises from the underlying quantum critical point separating the cohesive phase in which $\gamma_0= 0$ and the fractionalized phase in which $\gamma_0\neq0$. As we have illustrated, this phase transition also separates phases in which $\rho_n^{(0)}$ does and does not vanish. Together, these observations give further evidence that a transition between two types of quantum critical phases may explain the phenomenology in the overdoped cuprates, see also \cite{2009arXiv0912.2001H,2016Natur.531..210B,2019arXiv190908102P}.

As a final remark, we observe that in a Lifshitz quantum critical fractionalized phase, our result \eqref{rhonbehaviorfrac} implies $\rho_s\simeq \rho_s^{(0)}+\#T^{1-\theta/z}$. Setting $\theta=0$, the superfluid density displays a universal $T$-linear scaling for all $z>1$. A similar observation was reported by recent experiments in overdoped La$_{2-x}$Sr$_x$CuO$_4$ \cite{Bozovic1}. For $z=2$, the heat capacity will also receive a $T$-linear contribution. The value $z=2$ has appeared previously in theoretical models of high $T_c$ superconductors, see e.g. \cite{abanov2003quantum,Metlitski:2010vm,Patel:2014jfa}.

\begin{acknowledgments}
\emph{Acknowledgments. }We would like to thank Nigel Hussey, Catherine Pepin, and Steve Kivelson for useful discussions. In addition, we would like to thank Tomas Andrade and Richard Davison for initial collaboration at an early stage of this project. We would also like to thank Luca Delacr\'etaz for many discussions on superfluid hydrodynamics. We are grateful to Richard Davison, Sean Hartnoll, Chris Herzog and Jan Zaanen for helpful comments on a previous version of this manuscript. We would especially like to thank Jan Zaanen who first brought the results of \cite{Bozovic1} to our attention at the Aspen Center for Physics, where this work was initiated and which is supported by National Science Foundation grant PHY-1607611. This work was supported by the European Research Council (ERC) under the European Union's Horizon 2020 research and innovation programme (grant agreement No.758759).

\end{acknowledgments}


\appendix

\section{Appendix A: Details of the hydrodynamics \label{apphydro}}
Here, we go into more detail about the linearized hydrodynamic fluctuations used to derive Eq.~(6) of the main text, keeping the discussion self-contained. For completeness, we will write everything in a gauge invariant way, using the variable $\xi_\mu \equiv \partial_\mu\varphi-A_\mu$ where $A_\mu$ is a background gauge field.

As pointed out in the main text, the system is described by the following equations
\begin{align}
\label{hydroequations_one}
\partial_\mu T^{\mu\nu} &=F^{\mu\nu}j_\mu,\nonumber\\
\partial_\mu j^{\mu} &=0,\nonumber\\
u^\mu\xi_\mu + \mu &= 0,\nonumber\\
\partial_\mu(su^\mu) &= 0.
\end{align}
The first two equations express the local conservation laws: the first is the conservation of the fluid stress tensor and the second is the conservation of the $U(1)$ symmetry current. The last two equations are required by gauge invariance and thermodynamics: the third equation is a ``Josephson equation" which relates the time component of the background gauge field to the phase of the superfluid.  The final equation states that in equilibrium, the entropy density is conserved. 

Hydrodynamics states that these equations can be solved in terms of a derivative expansion of local thermodynamic variables and the fluid velocity. At non-dissipative order, we may write
\begin{align}
T^{\mu\nu} &= (\epsilon_n + P)u^\mu u^\nu + P\eta^{\mu\nu} + \frac{\rho_s}{\mu} \xi^\mu \xi^\nu\nonumber\\
j^\mu &= \rho_n u^\mu + \frac{\rho_s}{\mu} \xi^\mu.
\end{align}
There are no contributions from $F_{\mu\nu}$ since these are first order in derivatives of the thermodynamic variables. In writing these equations, we have explicitly chosen the entropy current to lie along $u^\mu$ since we expect only the normal component to transport entropy. This serves as a \emph{definition} of the normal component of the fluid. The total charge density is $\rho = \rho_n + \rho_s$ and the normal energy density, $\epsilon_n$, and pressure, $P$, satisfy the Smarr relation and Gibbs relations,
\begin{align}
\epsilon_n + P &= Ts+\rho_n\mu, \nonumber\\
dP = sdT + \rho d\mu &- \frac{\rho_s}{2\mu} d(\xi_\nu\xi^\nu + \mu^2)
\end{align}
Note that the true energy density, $\epsilon\equiv T^{00} = \epsilon_n+\mu\rho_s$.

We now look at fluctuations about equilibrium. Here it is useful to choose a frame, for instance one in which the normal and superfluid components are at rest. We emphasize that this choice does not affect the argument. We write $T=T^{(0)} +\delta T$, $\mu = \mu^{(0)} + \delta\mu $, $u^\mu= (1,\delta u^i)^\mu$, $\xi_0 = -\mu + \delta \xi_0$, $\xi_i = \delta \xi_i$. The fluctuations in the fluid stress tensor and current become. 
\begin{align}
\label{hydrofluctuations}
\delta T^{00} &= \delta \epsilon\nonumber\\
\delta T^{0i} &= (\mu\rho_n + sT)\delta u^i + \rho_s \delta\xi^i\nonumber\\
\delta T^{ij} &= \delta P \; \delta^{ij}\nonumber\\
\delta j^0 &= \delta \rho\nonumber\\
\delta j^i &= \rho_n \delta u^i + \frac{\rho_s}{\mu}\delta\xi^i\nonumber\\
\delta\xi_0 &=- \delta \mu
\end{align}
Hence, the set of equations (\ref{hydroequations_one}) now become,
\begin{align}
0&=\partial_0 \delta \epsilon + (\mu\rho_n + sT)\partial_i \delta u^i + \rho_s \partial_i\delta\xi^i \nonumber\\
\rho\;\delta F_{0i}&=(\mu\rho_n + sT)\partial_0\delta u^i + \rho_s \partial_0\delta\xi^i+\partial^i\delta P\nonumber\\
0&=\partial_0\delta\rho + \rho_n\partial_i\delta u^i +\frac{\rho_s}{\mu}\partial_i\delta\xi^i\nonumber\\
0&=\partial_i\delta\xi_0+ \partial_i\delta\mu\nonumber\\
\end{align}
Using the Gibbs relation, the second equation can be rewritten
\begin{align}
-\rho\delta F_{i0}=(\mu\rho_n + sT)\partial_0\delta u_i + \rho_s\partial_0\delta\xi_i + s\partial_i\delta T + (\rho_n+\rho_s)\partial_i\delta\mu.
\end{align}
Finally, if $s\to 0$ as $T\to 0$, consistency of these equations requires
\begin{align}
\rho_n\partial_0(\mu\delta u_i - \delta\xi_i) = 0.
\end{align}
In \cite{Leggett}, supposing independence of $\delta u_i$ and $\delta\xi_i$ requires  $\rho_n = 0$. As we show in the main text, $\partial_0(\mu\delta u_i - \delta \xi_i)$ generically vanishes. We can obtain this conclusion by consistently considering sources for the thermodynamic variables in the Hamiltonian \cite{Valle:2007xx}.

An external source deforms the Hamiltonian in linear response as
\begin{align}
\delta H(t) = -\sum_{A}\int d^2\vec{x}A(\vec{x},t)\delta s_A(\vec{x},t).
\end{align}
A small constant source applied in the infinite past and suddenly switched off at $t=0$,
\begin{align}
\delta s_A(\vec{x},t) = \delta s_A(\vec{x})e^{\varepsilon t}\theta(-t)
\end{align}
leads to a response in $A$ given by the matrix of static susceptibilities (Fourier transformed),
\begin{align}
\delta \langle A (\vec{k},t=0) = \sum_{B} \chi_{BA}(\vec{k}) \delta s_B(\vec{k}).
\end{align}
Here $\varepsilon\to 0^+$ leads to nice analyticity properties. Notably, if the source is a hydrodynamic variable, then $\chi_{AB} = \chi_{BA}$.

Define $v^i = \frac{1}{\mu}\xi^i$. The first law tells us that the Hamiltonian must contain a term
\begin{align}
H  \supset  \int d^2x \frac{\mu\rho_s}{2}\sum_i (v^i)^2. 
\end{align}
Fluctuations in $v^i$ due to a source $s_{v,i}$ are obtained via the Hamiltonian deformation,
\begin{align}
\delta H = -\int d^2 x \sum_{i} s_{v,i}v^i
\end{align}
which, at linear order, implies from \eqref{hydrofluctuations},
\begin{align}
\langle v^i\rangle = \frac{s_v^i}{\mu\rho_s} \Rightarrow \chi_{P^i v^i} \equiv \frac{\delta T^{i0}}{\delta s_v^{i}} = 1.
\end{align}
On the other hand, we must have
\begin{align}
\chi_{P^iv^i} = \chi_{v^iP^i} \equiv \frac{\delta v^i}{\delta u^i} = 1
\end{align}
so in fact
\begin{align}
s_v^i = \mu \rho_s(v^i - u^i).
\end{align}
In the absence of external sources, $s_v^i = 0 \Rightarrow v^i = u^i$ so that the superfluid and normal velocities are aligned and we have
\begin{align}
\mu \delta u_i = \xi_i
\end{align}
in linear response. We further note that within the context of linear response, the equations (\ref{hydrofluctuations}) also imply the electric conductivity (at non-dissipative order)
\begin{align}
\label{sigmahydrolowfreq}
\sigma(\omega) = \frac{i}{\omega}\left[\frac{\rho_n^2}{\mu\rho_n+sT} + \frac{\rho_s}{\mu}\right].
\end{align}
Since $\rho = \rho_n+\rho_s$, the pole in the imaginary frequency can be used to directly find $\rho_n$
\begin{align}
\label{impolesigmahydrolowfreq}
\lim_{\omega\to 0}\omega\text{Im}[\sigma(\omega)] = \frac{\rho}{\mu} - \left(\frac{sT}{\mu}\right)\frac{\rho_n}{\mu\rho_n+sT}.
\end{align}

Next, consider fluctuations with a spatially varying source with momentum $k^\mu = (0,\hat{k}^i)^\mu$. Fluctuations in the superfluid velocity are parallel to $\hat{k}$. Hence, if we look at transverse fluctuations, $\hat{\xi}$ does not contribute. In particular, the transverse momentum fluctuations obey
\begin{align}
\delta T^{0\perp} = (\mu\rho_n + sT)\delta u^\perp \Rightarrow \chi_{P^\perp P^\perp} = \mu\rho_n +sT
\end{align}
where $\chi_{P^\perp P^\perp}$ follows from a discussion similar to $\chi_{\hat v \hat P}$.  So far, we have omitted dissipative terms in our hydrodynamic discussion. They can be found, for instance, in \cite{Valle:2007xx, Herzog:2011ec}. Upon their inclusion, one finds a diffusion pole satisfying the Einstein relation
\begin{align}
\chi_{P^\perp P^\perp} = \frac{\eta}{D_\perp},
\end{align}
see for instance the discussion below Eq.~(4.21) of \cite{Valle:2007xx}. Here, $\eta$ is the shear viscosity which relaxes gradients in the transverse component of the normal fluid velocity and $D_\perp$ is the momentum diffusion constant which controls the rate of relaxation of the conserved transverse momentum. We also find two propagating modes, an adiabatic sound mode with $c_1^2 = (\partial P/\partial \epsilon)_{S,N}$, and second sound \cite{Valle:2007xx},
\begin{align}
c_2^2 = \left(\frac{s}{\rho}\right)^2 \frac{\rho_s}{(\partial[s/\rho]/\partial T)_\mu} \left(\frac{D_\perp}{\eta}\right)
\end{align}
That the second sound velocity and transverse diffusion/viscosity are related may be surprising because naively second sound is a longitudinal mode whereas shear diffusion is a transverse mode. However, as opposed to conventional sound which transports density fluctuations, second sound only transports heat while maintaining a constant local charge density (the normal and superfluid components are out of phase). This relaxation of transverse velocity gradients in the normal fluid is an effective bottleneck for superfluid sound that connects $D_\perp$ to $c_2$ and vice-versa. 

In addition to this observation, holography suggests that there exists a universal velocity, $v_{IR}^2$ in the infrared which provides a lower bound on diffusion, $(4\pi T)D_\perp \gtrsim v_{IR}^2$ \cite{Hartnoll:2014lpa}. In isotropic holographic models with translation invariance, $\eta= s/4\pi$ \cite{Kovtun:2004de}. For phases discussed in the main text that have $\rho_n \sim c_{IR}^{-2} (sT) \lesssim sT$ we find
\begin{align}
D_\perp \simeq c_{IR}^2/(4\pi T).
\end{align}
This agrees with \cite{Blake:2016wvh} where $v_B = c_{IR}$. On the other hand, for phases in which $\rho_n^{(0)}\neq 0$, we find
\begin{align}
D_{\perp} \simeq \frac{s}{4\pi \mu \rho_n^{(0)}}.
\end{align}

\section{Appendix B: Details of the holographic model\label{app:model}}
The holographic action 
\begin{align}
\label{holographicaction}
S = \int d^{d+2}x&\sqrt{-g}\Biggl[R-\frac{Z(\phi)}{4}F^2 - |D\eta|^2 - \frac{1}{2}(\partial\phi)^2- V(\phi,|\eta|) - \frac{Y(\phi)}{2}\sum_{i=1}^{d}(\partial\psi_i)^2\Biggr].
\end{align}
 gives the following field equations
\begin{align}
0=&G_{\mu\nu} + \frac{Z}{2}F_{\mu\lambda}F^{\lambda}_{\;\;\nu} - \frac{1}{2}\partial_\mu\phi\partial_\nu\phi- \frac{Y}{2}\partial_\mu \psi_I\partial_\nu\psi_I- (D_\mu\eta)(D_\nu\eta)^* \nonumber\\
&+ \frac{g_{\mu\nu}}{2}\left[\frac{1}{2}(\partial\psi)^2+|D\eta|^2+V+\frac{Z}{4}F^2+\frac{Y}{2}\left(\partial \psi_I\right)^2\right]\nonumber\\
0=&\frac{1}{\sqrt{-g}}\partial_\mu(\sqrt{-g}\partial^\mu\phi) - \frac{1}{4}\partial_\phi Z F^2 + \partial_\phi V+\frac{1}{2}\partial_\phi Y (\partial\psi_I)^2\nonumber\\
0=&\frac{1}{\sqrt{-g}}\partial_\mu(\sqrt{-g}ZF^{\mu\nu}) -2q^2 \eta^2A^\nu\nonumber\\
0=&\frac{1}{\sqrt{-g}}\partial_\mu(\sqrt{-g}g^{\mu\nu}D_\nu\eta)+ iq(D_\nu\eta)A^\nu - \partial_{\eta^*}V\nonumber\\
0=&\frac{1}{\sqrt{-g}}\partial_\mu[\sqrt{-g}g^{\mu\nu}(D_\nu\eta)^*]+ iq(D_\nu\eta)^*A^\nu - \partial_\eta V\nonumber\\
0=&\frac{1}{\sqrt{-g}}\partial_\mu\left(\sqrt{-g}Y\partial^\mu\psi_I\right)
\end{align}
where $G_{\mu\nu} = R_{\mu\nu} - \frac{1}{2}Rg_{\mu\nu}$ is the Einstein tensor.

We use the following ansatz for the metric and matter fields consistent with the staticity and rotational symmetry,
\begin{align}
ds^2 &= -D(r)dt^2 + B(r)dr^2 + C(r)(dx^2+dy^2),\nonumber\\
A &= A_t dt, \;\;\phi = \phi(r),\;\; |\eta| = \eta(r), \;\;\psi_I = mx^j\delta_{Ij}
\end{align}
which gives the equations of motion
\begin{align}
0&=\frac{d}{dr}\left[C\sqrt{\frac{D}{B}}\eta'\right]+q^2C\sqrt{\frac{B}{D}}A_t^2\eta - \frac{1}{2}C\sqrt{BD}\partial_\eta V\nonumber\\
0&=\frac{d}{dr}\left[C\sqrt{\frac{D}{B}}\phi'\right]+\partial_\phi Z\frac{C}{2\sqrt{BD}}(A_t')^2-m^2\sqrt{BD}\partial_\phi Y- C\sqrt{BD}\partial_\phi V\nonumber\\
0&=\frac{d}{dr}\left[Z\frac{C}{\sqrt{BD}}A_t'\right] -2q^2 C\sqrt{\frac{B}{D}}\eta^2A_t\nonumber\\
0&=\left(\frac{C}{\sqrt{BD}}D'\right)' + C\sqrt{BD}V - \frac{CZ(A_t')^2}{2\sqrt{BD}}-2q^2 C\sqrt{\frac{B}{D}}\eta^2A_t^2\nonumber\\
0&=(\frac{C'}{\sqrt{BCD}})' + \frac{1}{2}\sqrt{\frac{C}{BD}}(\phi')^2+ \sqrt{\frac{C}{BD}}(\eta')^2+q^2\frac{BC}{D^{3/2}}\eta^2A_t^2\nonumber\\
0&=(\phi')^2 +2(\eta')^2- \frac{C'}{C}\left(2\frac{D'}{D}+\frac{C'}{C}\right)- Z\frac{(A_t')^2}{D} - 2BV + 2q^2\frac{B}{D}\eta^2A_t^2
\end{align}
The scalar $\psi_I$ equations are trivially satisfied by our ansatz.

From these equations, we defined the flux from the black hole horizon as
\begin{align}
\rho_{in} \equiv -Z\frac{C}{\sqrt{BD}}A_t'\biggr|_{r=r_h}
\end{align}
where $r=r_h$ is the radial location of the black hole horizon. When $\lim_{T\to 0} \rho_{in} \equiv \rho_{in}^{(0)}= 0$, we are in a cohesive phase and when $\rho_{in}^{(0)}\neq 0$, we are in a fractionalized phase.

The equations can be combined to the simple equation
\begin{align}
\label{conservationequation}
\frac{d}{dr}\left[\frac{C^2}{\sqrt{BD}}\left(\frac{D}{C}\right)' - \frac{ZC}{\sqrt{BD}}A_t'A_t\right]=m^2\sqrt{BD}Y.
\end{align}
which is seen to be conserved for $m=0$. The term inside the square brackets gives $(-sT)$ when evaluated on the horizon.

In the UV ($r \to 0$), we require that the metric functions and matter fields have an expansion,
\begin{align}
\phi(r) &= \mu - \rho\; r +...\nonumber\\
\phi(r) &= \phi_s\; r + \phi^{(2)}r^2+...\nonumber\\
\eta(r) &= \eta^{(1)}r+\frac{\langle O_2\rangle}{\sqrt{2}}r^2+...\nonumber\\
D(r) &= r^{-2} - \frac{2m^2+\psi_1^2}{8} - \frac{\langle \epsilon \rangle}{3} \;r+...\nonumber\\
C(r) &= r^{-2} + \frac{2m^2-\psi_1^2}{8} - \frac{\langle P \rangle}{3} \;r+...\nonumber\\
B(r) &= r^{-2}+...
\end{align}
Here, $\eta^{(1)}$ is a source for the complex scalar. When $\eta^{(1)}=0$ and $\langle O_2\rangle \neq 0$, the $U(1)$ symmetry is spontaneously broken. The factor of $\sqrt{2}$ is a normalization convention \cite{Hartnoll:2008kx}. Next , $\phi_s$ is a source term for the dilaton which we can use to vary the condensation temperature $T_c$ and drive a phase transition between fractionalized and cohesive phases. Notably, sourcing $\phi$ breaks conformal invariance as reflected in the trace of the stress tensor.
When the asymptotics are inserted into the conservation equation \eqref{conservationequation}, we derive the Smarr relation,
\begin{align}
\epsilon+P = \mu\rho + Ts.
\end{align}
Inserting this expansion into the equations of motion gives
\begin{align}
P = \frac{\epsilon}{2} + \left(\frac{\eta^{(1)}\langle O_2\rangle}{\sqrt{2}} + \frac{1}{2}\phi_s\psi^{(2)}\right).
\end{align}
Variation of this pressure shows that $\eta^{(1)}=0$ means the superfluid velocity is not sourced. The $\epsilon$ that appears here is the true energy density, rather than the normal energy density $\epsilon_n$. The two are related by $\epsilon=\epsilon_n+\mu \rho_s$.

\section{Appendix C: Holographic computation of the conductivity and of the normal and superfluid densities \label{app:conductivity}}

The conductivity is obtained by sourcing a fluctuating spatial component of the gauge field, $\delta a_x = a_x(r)e^{-i\omega t}$. For $m=0$, this requires sourcing $\delta g_{tx} =  g_{tx}(r)e^{-i\omega t}$ and for $m\neq 0$ we must also source $\delta g_{xr} = g_{xr}(r)e^{-i\omega t}$ and a fluctuation in $\psi_x = \psi_x + \delta \psi_x e^{-i\omega t}$, see for instance \cite{Andrade:2013gsa}. For ease of reading, we will only write the translation invariant equations. Defining $F=Z\sqrt{\frac{D}{B}}$,
\begin{align}
\label{axequation}
0&=\frac{d}{dr}\left[Fa_x'\right]- \frac{Z}{F}\left(2q^2D\eta^2-\omega^2Z+\frac{F^2(A_t')^2}{D}\right)a_x\\
0&=\frac{d}{dr}\left[\frac{g_{tx}}{C}\right] + \frac{Z}{C}a_xA_t'
\label{constraintequation}
\end{align}

In the UV, the fluctuations behave as, 
\begin{align}
a_x(r) = a_x^{(0)} + a_x^{(1)} r +...\nonumber\\
g_{tx}(r) = r^{-2}g_{tx}^{(0)} - \frac{\langle T_{tx} \rangle}{3} r +...
\end{align}
The applied electric field is $F_{xt} = i\omega a_x^{(0)}$ and if we do not apply a temperature gradient $g_{tx}^{(0)}=0$. The frequency-dependent conductivity is, following the holographic renormalization procedure \cite{Hartnoll:2008kx,Skenderis:2002wp},
\begin{align}
\sigma(\omega) = \frac{a_x^{(1)}}{i\omega a_x^{(0)}}.
\end{align}

We are interested in extracting the normal and superfluid charge densities, which can be read off from the low frequency behavior of the ac conductivity through \eqref{sigmahydrolowfreq} and \eqref{impolesigmahydrolowfreq}. As we now explain, they can be computed more simply by solving the $\omega=0$ limit of \eqref{axequation}.

This equation has two independent solutions, one regular at the horizon and another which is singular there, given by the Wronskian. At low frequencies, a matching argument shows that the singular solution does not contribute to the imaginary part of the conductivity \cite{Davison:2015taa}:
\begin{align}
\label{axomega0sol}
a_x(r)=a^{(reg)}_x(r)+i\omega Z(\phi(r_h))\left(a^{(reg)}_x(r_h)\right)^2 \frac{a^{(sing)}_x(r)}{a_x^{(0)}}\,.
\end{align}
Thus, it is enough to compute $a^{(reg)}_x$ to read off the weight of the imaginary pole of the conductivity, which together with the relation $\rho=\rho_n+\rho_s$, gives access to both the normal and superfluid densities. 

At zero temperature, the fluctuation equations reduce to imply that $a^{(reg)}_x = \frac{a_x^{(0)}}{\mu}A_t$, where $A_t$ is the background solution for the gauge field. Plugging into \eqref{axomega0sol}, this gives
\begin{align}
\lim_{\omega\to0}\text{Im}[\sigma(\omega)] = \frac{\rho}{\mu}
\end{align}
as expected.

Furthermore, the constraint equation \eqref{constraintequation} reads
\begin{align}
\langle T_{tx} \rangle = - \rho a_x^{(0)}
\end{align}
Gauge invariance allows us to trade $a_x^{(0)}\leftrightarrow \xi_x$ and inspecting \eqref{hydrofluctuations}, we find that this equation requires $\delta u^x = \xi^x/\mu$ at $T=0$ which is consistent with our discussion of the static susceptibility matrix.

\section{Appendix D: $\phi_s/\mu$ as a proxy for doping \label{app:doping}}
In the main text, we stated that $\phi_s/\mu$ could be used as a proxy for doping. We justify this in the following way. By varying $\phi_s/\mu$, a UV quantity, we approach two quantum critical points in the IR at values $\phi_s^{*}$ and $\phi_s^c$, while we simultaneously change the critical temperature for superfluidity, as indicated in Figure 3 of the main text. In addition to this behavior, increasing $\phi_s/\mu$ simultaneously increases the total charge density at zero temperature, $\rho_T^{(0)}$ and enhances the scattering rate $\Gamma = m^2 s Y(r_h)/(4\pi[\mu \rho_n + sT])$. We illustrate this in the following plots. This qualitatively behaves analogous to doping in the cuprates.

\begin{figure}[h]
\centering
\includegraphics[width=.3\columnwidth]{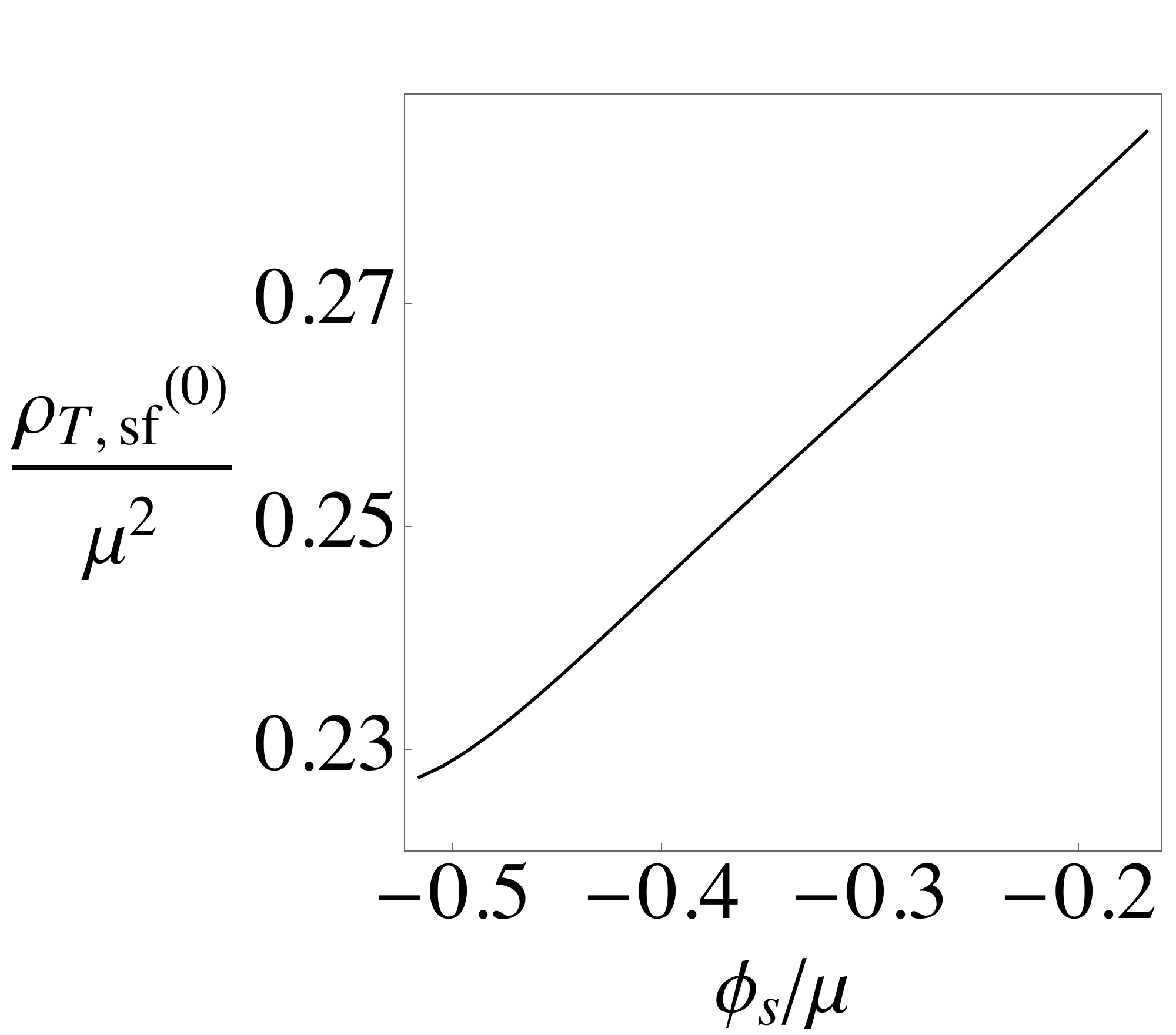}
\includegraphics[width=.3\columnwidth]{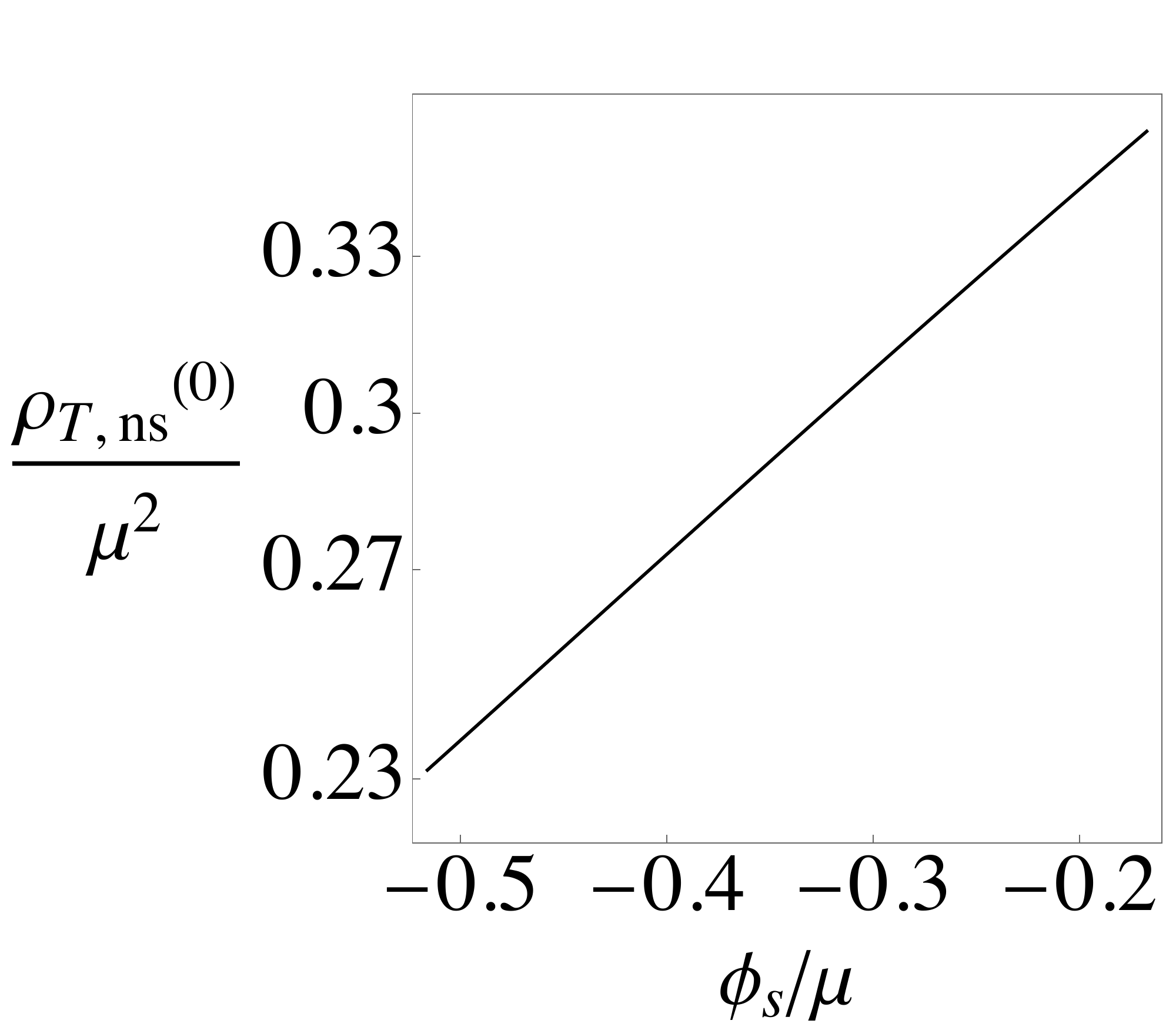}
\includegraphics[width=.3\columnwidth]{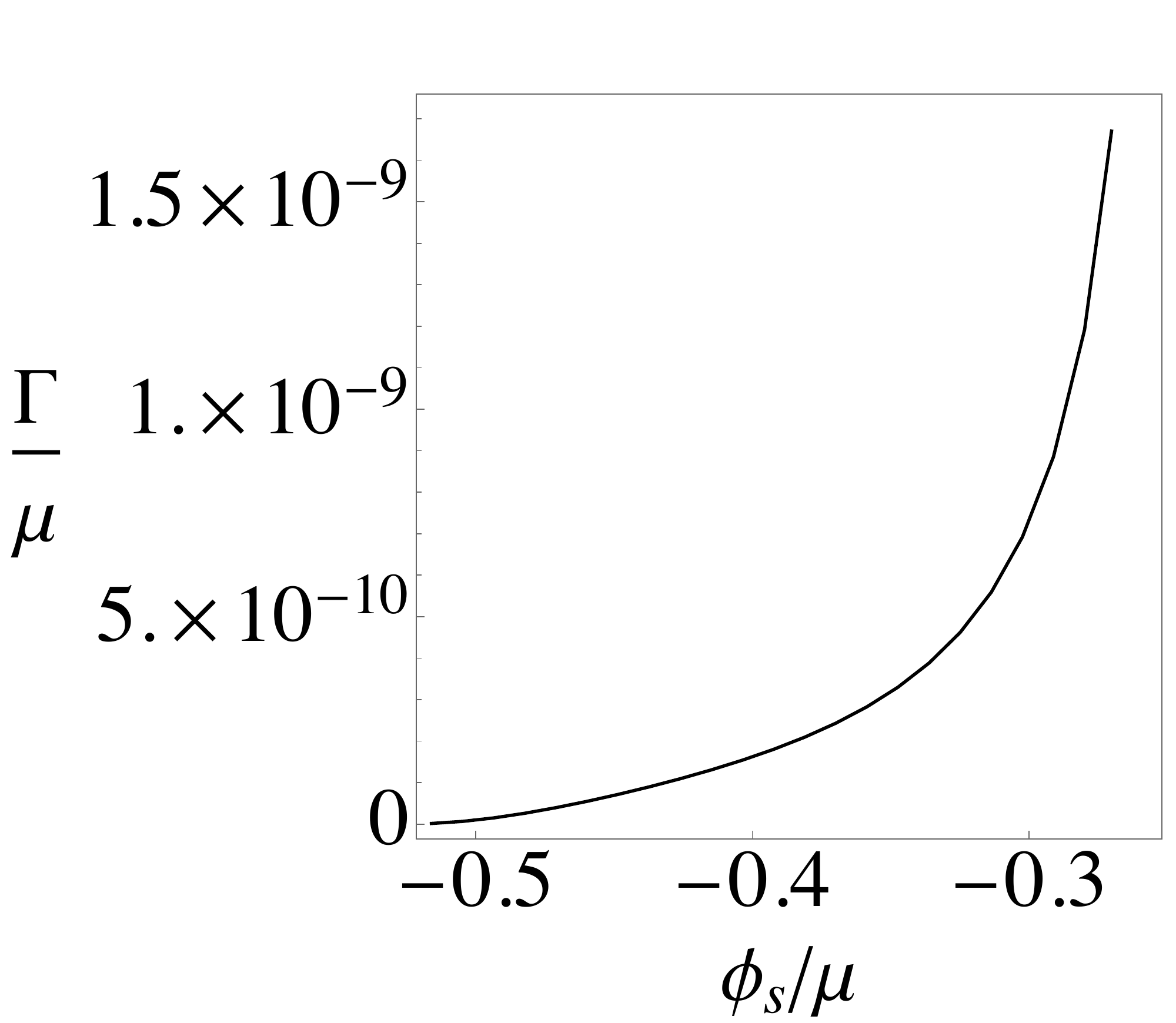}
\caption{The total charge density as a function of doping with a superfluid (left) and without a superfluid (center). The scattering rate as a function of doping (right).}
\end{figure}

\bibliography{local}

\end{document}